\documentclass[aps,prl,floatfix,twocolumn]{revtex4}
\usepackage{graphicx}

\input{tcilatex}
\begin{document}

\title{Interference of Bose-Einstein condensates: quantum non-local effects}
\author{W. J. Mullin $^{a}$ and F. Lalo\"{e} $^{b}$}
\affiliation{$^{a}$Department of Physics, University of Massachusetts, Amherst,
Massachusetts 01003 USA\\
$^{b}$ Laboratoire Kastler Brossel, ENS, UPMC, CNRS; 24 rue Lhomond, 75005
Paris, France}

\begin{abstract}
Quantum systems in Fock states do not have a phase.\ When two or more
Bose-Einstein condensates are sent into interferometers, they nevertheless
acquire a relative phase under the effect of quantum measurements.\ The
usual explanation relies on spontaneous symmetry breaking, where phases are
ascribed to all condensates and treated as unknown classical quantities.\
However, this image is not always sufficient: when all particles are
measured, quantum mechanics predicts probabilities that are sometimes in
contradiction with it, as illustrated by quantum violations of local
realism.\ In this letter, we show that interferometers can be used to
demonstrate a large variety of violations with an arbitrarily large number
of particles.\ With two independent condensates, we find violations of the
BCHSH inequalities, as well as new $N$-body Hardy impossibilities.\ With
three condensates, we obtain new GHZ (Greenberger, Horne and Zeilinger) type
contradictions.
\end{abstract}

\maketitle

Gaseous Bose-Einstein condensates (BEC) can be used as sources to perform
experiments with atomic interferometers \cite{Simsarian}.\ With a single
condensate, the interference effects depend on the difference of the
accumulated phases along the arms of the interferometer. With two or more
BEC's, their relative phase introduces new physics into the problem.\ The
usual view is that, when spontaneous symmetry breaking takes place at the
Bose-Einstein transition, each condensates acquires a phase, with a
completely random value. The outcome of a given experiment can then be
obtained by assuming the existence of this initial classical phase; for an
ensemble of realizations, an average over all of its possible values is
necessary.\ Spectacular experiments with alkali atoms originating from two
independent BEC\ have confirmed this view \cite{WK}.\ Long before, Anderson 
\cite{Anderson} had proposed a thought experiment raising the famous
question \textquotedblleft Do superfluids that have never seen each other
have a well-defined relative phase?\textquotedblright\ The question is not
trivial since, in quantum mechanics, the Bose-Einstein condensates of
superfluids are naturally described by Fock states, for which the phase is
completely undetermined. Nevertheless, various authors \cite{Java, CD} have
shown that repeated quantum measurements of the relative phase of two Fock
states make a well-defined value emerge spontaneously with a random value 
\cite{Hall, LM}. Then, considering that the phase appears under the effect
of spontaneous symmetry breaking, when the BEC's are formed, or later, under
the effect of measurements, seems to be only as a matter of taste.

But a closer examination of the problem shows that this is not always true 
\cite{LS}: situations do exist where the two points of view are not
equivalent, and even where the predictions of quantum mechanics for an
ensemble of measurements are at variance with those obtained from an average
over a phase \cite{LM}.\ This is not so surprising after all: the idea of a
pre-existing phase is very similar to the notion of \textquotedblleft
elements of reality\textquotedblright\ \cite{FL} introduced by Einstein,
Podolsky and Rosen \cite{EPR}--for a double Fock state, the relative phase
is nothing but what is often called a \textquotedblleft hidden
variable\textquotedblright --and we know that this idea combined with
locality leads to the Bell theorem \cite{Bell} and to contradictions with
quantum mechanics.\ It is then natural that the notion of classical phase
should also lead to Bell-type inequalities and to similar contradictions.

Such contradictions were indeed predicted in \cite{LM}, but in conditions
that seemed difficult to reach experimentally: precise spin measurements in $%
N$ separate regions of space were required ($N$ is the total number of
particles), and the numbers of results at each end of the experiment were
fixed. Here we consider more realistic situations where spinless particles
enter interferometers, and where the only requirement is to measure in which
arm they leave them; this is accessible by laser atomic fluorescence
(repeated measurements are possible in a quantum non-demolition scheme).\
Moreover, the number of results in each region may fluctuate freely.\ We
study various situations involving two or three BEC's, used as sources for
interferometers, and show that quantum mechanics predicts violations of the
BCHSH inequalities \cite{CHSH}, of the GHZ contradictions \cite{GHZ,GHZ2} as
well as of the Hardy impossibilities \cite{H-1,H-2}. Fock-state condensates
appear as remarkably versatile, able to create violations that usually
require elaborate entangled wave functions, and produce \emph{new} $N$-body
violations.

We first study an interferometer with a double Fock-state representing the
sources, as shown in Fig.1: a condensate containing $N_{\alpha }$ particles
reaches a semi-reflecting plate and is split into two coherent components $u$
and $v$; similarly, another condensate containing $N_{\beta }$ particles
reaches another semi-reflecting plate and is split into two components $w$
and $t$.\ We assume that, in two remote regions of space $D_{A}$ and $D_{B}$%
, two experimenters Alice and Bob make measurements with semi-reflecting
plates, recombining components $v$ and $w$ for the former, $u$ and $t$ for
the latter; before the plates, they insert devices providing a phase shift, $%
\zeta $ for Alice, $\theta $ for Bob.\ We call $m_{1}$ and $m_{2}$ the
number of particles that Alice detects in output $1$ and $2$ respectively,
and $m_{3}$ and $m_{4}$ the similar quantities for Bob. Ref. \cite{YS1}
gives a study of a two-particle Bell inequality with this interferometer. We
now calculate the probability $P(m_{1},m_{2},m_{3},m_{4})$ of such events.

\begin{figure}[h]
\centering \includegraphics[width=2.5in]{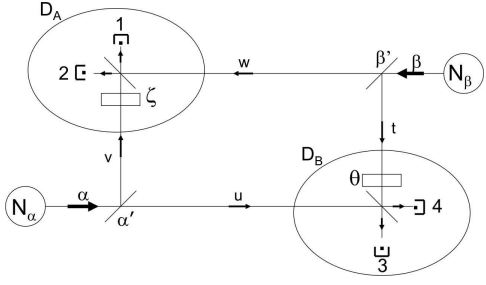}
\caption{Two independent condensates (populations $N_{\protect\alpha }$ and $%
N_{\protect\beta }$) are split into two coherent components, and then enter
interferometers in two remote places $D_{A}$ and $D_{B}$; the quantum
results strongly violate the BCHSH\ local realist inequalities.}
\label{fig1}
\end{figure}

The destruction operators $a_{1}\cdots a_{4}$ associated with the output
modes can be written in terms of the mode operators at the sources $%
a_{\alpha },a_{\beta },a_{\alpha ^{\prime }}$ and $a_{\beta ^{\prime }}$ by
tracing back from the detectors to the sources, with a phase shift of $\pi
/2 $ at each reflection, $\zeta $ or $\theta $ at the shifters, and a $1/%
\sqrt{2}$ at each beam splitter. This gives the projections of the two
different source modes onto each detector mode 
\begin{eqnarray}
a_{1} &=&\frac{1}{2}\left[ ie^{i\zeta }a_{\alpha }+ia_{\beta }\right] 
\nonumber \\
a_{2} &=&\frac{1}{2}\left[ -e^{i\zeta }a_{\alpha }+a_{\beta }\right] 
\nonumber \\
a_{3} &=&\frac{1}{2}\left[ ia_{\alpha }+ie^{i\theta }a_{\beta }\right] 
\nonumber \\
a_{4} &=&\frac{1}{2}\left[ a_{\alpha }-e^{i\theta }a_{\beta }\right]
\label{awm}
\end{eqnarray}%
where we have eliminated $a_{\alpha ^{\prime }}$ and $a_{\beta ^{\prime }}$,
which do not contribute; in short, we write these equations as $%
a_{i}=u_{i\alpha }a_{\alpha }+u_{i\beta }a_{\beta }$. The source state is 
\begin{equation}
\left\vert \Phi \right\rangle ~=\frac{1}{\sqrt{N_{\alpha }!N_{\beta }!}}%
a_{\alpha }^{\dagger N_{\alpha }}a_{\beta }^{\dagger N_{\beta }}\left\vert 
\text{0}\right\rangle  \label{DFS}
\end{equation}%
where $a_{\alpha }^{\dagger }$ and $a_{\beta }^{\dagger }$ are creation
operators and $\left\vert \text{0}\right\rangle $ is the vacuum.\ The
amplitude for the system crossing all beam splitters with $m_{1}\cdots m_{4}$
particles at the detectors is:

\begin{widetext}
\begin{equation}
C_{m_{1},..,m_{4}}=\left\langle m_{1},m_{2},m_{3},m_{4}\right\vert \left.
\Phi \right\rangle ~=\left\langle \text{0}\right\vert \frac{%
a_{1}^{m_{1}}\cdots a_{4}^{m_{4}}}{\sqrt{m_{1}!\cdots m_{4}!}}\frac{%
a_{\alpha }^{\dagger N_{\alpha }}a_{\beta }^{\dagger N_{\beta }}}{\sqrt{%
N_{\alpha }!N_{\beta }!}}\left\vert \text{0}\right\rangle  \label{C}
\end{equation}%
We substitute (\ref{awm}) into this expression, and make binomial expansions
of the sums $\left( u_{i\alpha }a_{\alpha }+u_{i\beta }a_{\beta }\right)
^{m_{i}}$ to find 
\begin{equation}
C_{m_{1},..,m_{4}}=\frac{1}{\sqrt{N_{\alpha }!N_{\beta }!}\sqrt{m_{1}!\cdots
m_{4}!}}\left( \prod_{i=1}^{4}\sum \frac{m_{i}!}{p_{\alpha i}!p_{\beta i}!}%
\left( u_{i\alpha }\right) ^{p_{\alpha i}}\left( u_{i\beta }\right)
^{p_{\beta i}}\right) \left\langle 0\right\vert a_{\alpha }^{p_{\alpha
1}+\cdots +p_{\alpha 4}}a_{\beta }^{p_{\beta 1}+\cdots +p_{\beta
4}}a_{\alpha }^{\dagger N_{\alpha }}a_{\beta }^{\dagger N_{\beta
}}\left\vert 0\right\rangle  \label{C-bis}
\end{equation}%
where the $\sum $ is a summation over $p_{\alpha i}$ and $p_{\beta i}$ with $%
p_{\alpha i}+p_{\beta i}=m_{i}$.\ The evaluation of the expectation value of
the operator product gives $N_{\alpha }!N_{\beta }!\delta _{N_{\alpha
},p_{\alpha 1}+\cdots p_{\alpha 4}~}\delta _{N_{\beta },~p_{\beta 1}+\cdots
p_{\beta 4}}$, which can be included by inserting the relations $\delta
_{N_{\gamma },p_{\alpha 1}+\cdots p_{\alpha 4}}=\int_{-\pi }^{\pi }\frac{%
d\lambda _{\gamma }}{2\pi }~e^{i(p_{\gamma 1}+\cdots p_{\gamma 4}-N_{\gamma
})\lambda _{\gamma }}$ with $\gamma =\alpha ,\beta $. If the total number of
measurements $M=\sum_{i}m_{i}$ is equal to the total number of particles $%
N=N_{\alpha }+N_{\beta }$, the probability of obtaining result $%
(m_{1},m_{2},m_{3},m_{4})$ is 
\begin{equation}
\mathcal{P}(m_{1},m_{2},m_{3},m_{4})=\left\vert C_{m_{1},\cdots
m_{4}}\right\vert ^{2}=\frac{N_{\alpha }!N_{\beta }!}{m_{1}!\cdots m_{4}!}%
\int d\tau \int d\tau ^{\prime }e^{-i\left[ N_{\alpha }(\lambda _{\alpha
}-\lambda _{\alpha }^{\prime })+N_{\beta }(\lambda _{\beta }-\lambda _{\beta
}^{\prime })\right] }\prod_{i=1}^{4}\left[ \Omega _{i}^{\prime \ast }\Omega
_{i}\right] ^{m_{i}}  \label{Prob}
\end{equation}%
with $d\tau $ representing integration over $\lambda _{\alpha }$ and $%
\lambda _{\beta }$ and $d\tau ^{\prime }$ over the $\lambda ^{\prime }$'s,
and $\Omega _{i}(\lambda _{a},\lambda _{b})=\left( u_{i\alpha }e^{i\lambda
_{\alpha }}+u_{i\beta }e^{i\lambda _{\beta }}\right) ,$ $\Omega _{i}^{\prime
}=\Omega _{i}(\lambda _{a}^{\prime },\lambda _{b}^{\prime })$. This
expression simplifies with the integration variables $\lambda _{\pm
}=[(\lambda _{\alpha }+\lambda _{\alpha }^{\prime })\pm (\lambda _{\beta
}+\lambda _{\beta }^{\prime })]/2;$ $\Lambda _{\pm }=[(\lambda _{\alpha
}-\lambda _{\alpha }^{\prime })\pm (\lambda _{\beta }-\lambda _{\beta
}^{\prime })]/2$, since two integrations then become trivial and disappear.\
Using $\Lambda _{-}$ parity, we then obtain 
\begin{equation}
\mathcal{P}(m_{1},m_{2},m_{3},m_{4})\sim \int_{-\pi }^{\pi }\frac{d\Lambda
_{-}}{2\pi }\cos (N_{\alpha }-N_{\beta })\Lambda _{-}\int_{-\pi }^{\pi }%
\frac{d\lambda _{-}}{2\pi }\prod_{i=1}^{4}\left[ \cos \Lambda _{-}+\eta
_{i}\cos \left( \lambda _{-}-\varphi _{i}\right) \right] ^{m_{i}}
\label{OldProb}
\end{equation}%
where $\eta _{1}=\eta _{3}=1$; $\eta _{2}=\eta _{4}=-1$; $\varphi
_{1}=\varphi _{2}=-\varsigma $; $\varphi _{3}=\varphi _{4}=\theta $. 
\end{widetext}

When $N=2$ it is easy to show that $\mathcal{P}(0,1,0,1)$ is equal to $\frac{%
1}{4}\cos ^{2}\left( \frac{\varsigma +\theta }{2}\right) $, in agreement
with Ref. \cite{YS1}. For any $N$, we recover the same form of the
probability as for two interfering spinor condensates \cite{LM}; detectors 1
and 3 in Fig. 1 correspond to results $\eta =+1$ for spin measurements, 2
and 4 to $\eta =-1$ results. Nevertheless, instead of assuming that Alice
and Bob measure a fixed number of spins, here the number of particles they
detect can fluctuate freely, which changes the averages.

If in (\ref{OldProb}) we set $\Lambda _{-}=0$, we obtain the predictions of
a pre-existing phase $\lambda _{-}$, with a product of phase-dependent local
probabilities $\frac{1}{2}\left[ 1+\eta _{i}\cos \left( \lambda _{-}-\varphi
_{i}\right) \right] $ summed over all possible values of $\lambda _{-}$.
But, when $\Lambda _{-}$ varies, negative \textquotedblleft
probabilities\textquotedblright\ appear in the integrand, introducing
limitations to the notion of classical phase and the possibility of
violations of local realism. This can happen only if all particles are
measured: if $M<N$ particles are detected, summing over the unobserved
results amounts to setting $\eta _{i}=0$ in the corresponding factors of (%
\ref{OldProb}), so that a factor $\left[ \cos \Lambda _{-}\right] ^{N-M}$
appears in the formula, peaked at $\Lambda _{-}=0$.\ Quantum violations then
disappear, while the notion of relative phase re-appears.

Counting $\eta =+1$ and $-1$ values as above, we can define their product at
Alice's location as a quantity $\mathcal{A}=\pm 1$, and at Bob's location
their product as $\mathcal{B}=\pm 1$; we then have two functions to which
the BCHSH theorem can be applied. The quantum average of their product is: 
\begin{equation}
\left\langle \mathcal{AB}\right\rangle =\sum_{m_{1}\cdots
m_{4}}(-1)^{m_{2}+m_{4}}~\mathcal{P}(m_{1},m_{2},m_{3},m_{4})  \label{CAB}
\end{equation}%
The $m_{i}$ sums can be done, leading to an exponential of a sum of three
terms that can be re-expanded in three series. If $M=N$, the four $\tau $
integrals are easy, resulting in Kronecker $\delta $'s that collapse the
sums to a single term: 
\begin{equation}
\left\langle \mathcal{AB}\right\rangle =\left[ \cos \left( \frac{\zeta
+\theta }{2}\right) \right] ^{N}\delta _{N_{\alpha },N_{\beta }}  \label{AB}
\end{equation}

Now, the CHSH inequality tells us that: 
\begin{equation}
\left\langle \mathcal{AB}\right\rangle +\left\langle \mathcal{AB}^{\prime
}\right\rangle +\left\langle \mathcal{A}^{\prime }\mathcal{B}\right\rangle
-\left\langle \mathcal{A}^{\prime }\mathcal{B}^{\prime }\right\rangle \leq 2
\label{AB-bis}
\end{equation}%
where letters with and without primes imply measurements at differing
angles. Alice's measurement angle is taken for convenience as $\phi
_{a}=\varsigma /2$ and Bob's as $\phi _{b}=-\theta /2$. We define $E(\phi
_{a}-\phi _{b})=\cos ^{N}\left( \phi _{a}-\phi _{b}\right) $, set $\phi
_{a}-\phi _{b}=\phi _{b}-\phi _{a^{\prime }}=\phi _{b^{\prime }}-\phi
_{a}=\xi $ and $\phi _{b^{\prime }}-\phi _{a^{\prime }}=3\xi $, and maximize 
$Q=3E(\xi )-E(3\xi )$ to find the greatest violation of the inequality for
each $N$. For $N=2$ we find $Q_{\max }=2.41$ in agreement with Ref. \cite%
{YS1}; for $N=4,$ $Q_{\max }=2.36$; and for $N\rightarrow \infty ,$ $Q_{\max
}\rightarrow 2.32.$ The system continues to violate local realism for
arbitrarily large condensates. Note that every source particle must be
measured, otherwise no violation is found \cite{LM}.

Consider next the arrangement of Fig. 2 with three Fock condensate sources
and three detector pairs; it will allow GHZ contradictions. A similar device
was discussed in Ref. \cite{YS2}, but with only one particle per source;
Ref. \cite{GHZ2} also considered measuring spinless particles in an
interferometer. We proceed as above to find a probability $P(m_{1}\cdots
m_{6})$. 
\begin{figure}[h]
\centering \includegraphics[width=2.5in]{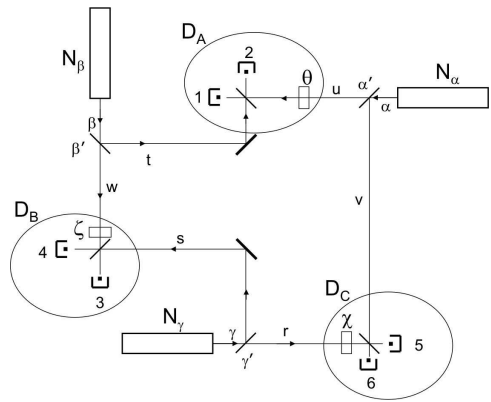}
\caption{Each of 3 condensates is split in 2 parts, which then enter
interferometers in 3 remote places $D_{A}$, $D_{B}$ and $D_{C}$; the quantum
results show GHZ contradictions with local realism.}
\label{fig2}
\end{figure}

For obtaining GHZ contradictions, we consider the case $N_{\alpha }=N_{\beta
}=N_{\gamma }=N/3$ and the events where all three detectors receive the same
number of particles. Because we are considering a limited number of cases,
the normalization is now different, and we must compute%
\begin{equation}
\begin{array}{l}
\mathcal{N}\equiv \sum_{m_{1}\cdots m_{6}}\delta _{m_{1}+m_{2},N/3}~\times
~\delta _{m_{3}+m_{4},N/3} \\ 
\multicolumn{1}{r}{\times \delta _{m_{5}+m_{6},N/3}~\times \mathcal{P}%
(m_{1}\cdots m_{6})}%
\end{array}
\label{N}
\end{equation}%
In order to limit the sum over the $m_{i}$'s to these cases, we introduce
three more integrals over $\rho _{A},$ $\rho _{B},$ and $\rho _{C}$ of the
form $\delta _{m_{1}+m_{2},N/3}=\int_{-\pi }^{\pi }\frac{d\rho _{A}}{2\pi }%
e^{i(m_{1}+m_{2}-N/3)\rho _{A}}$. To find the average $\left\langle \mathcal{%
ABC}\right\rangle $ for measurements done by Alice, Bob, and Carole we
introduce $(-1)^{m_{2}+m_{4}+m_{6}}$ into the sums over the $m_{i}$ and
perform them in the same way. Dividing by the normalization gives us 
\begin{equation}
\left\langle \mathcal{ABC}\right\rangle =\frac{\sum_{q}\left( \frac{N/3!}{%
(N/3-q)!q!}~\right) ^{3}e^{i(\varsigma +\theta +\chi )(N/3-2q)}}{%
\sum_{q}\left( \frac{N/3!}{(N/3-q)!q!}~\right) ^{3}}  \label{ABC}
\end{equation}

In the usual GHZ case $N=3$, we find $\left\langle \mathcal{ABC}%
\right\rangle =\cos (\varsigma +\theta +\chi )$, and perfect correlations if
the sum of angles is $\pi $. Local realism then gives $\left\langle \mathcal{%
ABC}\right\rangle $ $=A(\varsigma )B(\theta )C(\chi )$ and%
\begin{equation}
\begin{array}{l}
A(\pi /2)~B(\pi /2)~C(0)=-1 \\ 
A(\pi /2)~B(0)~C(\pi /2)=-1 \\ 
A(0)~B(\pi /2)~C(\pi /2)=-1%
\end{array}
\label{ABC-bis}
\end{equation}%
But then we must have $A(0)B(0)C(0)=-1,$ while quantum mechanics gives $+1$.
For larger $N$ we also obtain contradictions, for instance when $N=9$, where 
\begin{equation}
\left\langle \mathcal{ABC}\right\rangle =\frac{1}{28}\left[ 27\cos
(\varsigma +\theta +\chi )+\cos 3(\varsigma +\theta +\chi )\right]
\label{ABC-ter}
\end{equation}%
Since both cosines change sign when the angles increase by $\pi $, the above
argument remains unchanged and, again, leads to complete sign contradiction.
Actually, any time $N/3$ is odd, we get a similar result for arbitrary $N.$

Hardy impossibilities are treated by use of the interferometer shown in
Fig.~3 based on the one discussed in Ref.~\cite{H-1} for $N=2$. 
\begin{figure}[h]
\centering \includegraphics[width=2.5in]{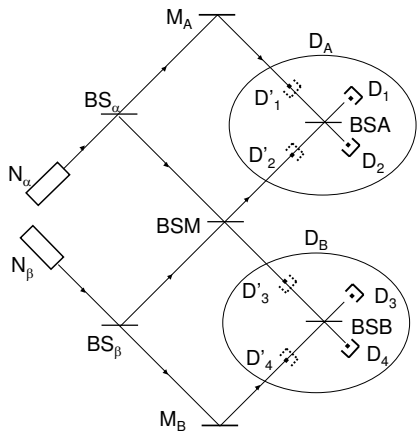}
\caption{An interferometer with particle sources $\protect\alpha $ and $%
\protect\beta $, beamsplitters BS and mirrors M. In both detection regions,
the detectors at $D_{i}$ may be replaced by the $D_{i}^{\prime }$, placed
before the beam splitters. For appropriate path lengths and reflectivities
of the beam splitters, quantum mechanics predicts the existence of events
that are forbidden by local realism.}
\label{Hardy}
\end{figure}
The heart of the system is the beam splitter at the center; due to Bose
interference it has the property that, if an equal number of particles
approaches each side, then an even number emerges from each side. The
detection beam splitters BSA and BSB are set to have a transmission
probability of 1/3 and the path differences are such that, by destructive
interference, no particle reaches $D_{2}$ if only source $N_{\alpha }$ is
used; similarly, no particle reaches $D_{3}$ if $N_{\beta }$ alone is used.
Alice can use either the detectors $D_{1,2}$ after her beam splitter, or $%
D_{1,2}^{\prime }$ before; Bob can choose either $D_{3,4}$, or $%
D_{3,4}^{\prime }$. This gives $4$ arrangements of experiments: $DD$, $%
DD^{\prime }$, $D^{\prime }D$, or $D^{\prime }D^{\prime }$, with probability
amplitudes $C_{XY}(m_{1},m_{2};m_{3},m_{4})$, where $XY$ is any of these $4$
arrangements and the $m$ values are the numbers of particles detected at
each counter.

Assume for instance that $N=6$ and that both experimenters observe $3$
particles. The probability amplitude $C_{D^{\prime }D^{\prime }}(0,3;3,0)$
vanishes because of the beam splitter rule. The destructive interference
effect at BSA and BSB lead to $C_{DD^{\prime }}(0,3;1,2)=C_{DD^{\prime
}}(0,3;2,1)=C_{DD^{\prime }}(0,3;0,3)=0$; but $C_{DD^{\prime }}(0,3;3,0)\neq
0$.\ Thus, if Alice observes $3$ particles at $D_{2}$, when Bob uses the
primed detectors he observes with certainty $3$ particles at $D_{3}^{\prime
} $; similarly, if Bob has seen 3 particles in $D_{3}$, in the $D^{\prime }D$
configuration Alice must see 3 in $D_{2}^{\prime }$ .

If both do unprimed experiments, we find $C_{DD}(0,3;3,0)=1/216$, which
shows that events exist where $3$ particles are detected at both detectors $%
D_{2}$ and $D_{3}$. In any of these events, if Bob had at the last instant
changed to the primed detectors, he would surely have obtained three
particles in $D_{3}^{\prime }$, because of the certainty mentioned above; if
Alice had changed detectors instead of Bob, she would have obtained $3$
particles in $D_{2}^{\prime }$. Now, had both changed their minds after the
emission and chosen the primed arrangement, local realism implies that they
would have found $3$ particles in each $D_{2}^{\prime }$ and $D_{3}^{\prime }
$: such events must exist.\ But its quantum probability is exactly zero, in
complete contradiction. This argument can be generalized to all cases of odd 
$N/2$ emitted from each source.

In conclusion, we think that the answer to Anderson's question is:
\textquotedblleft In most cases, this view leads to correct quantum
predictions, but not always.\textquotedblright\ It is sufficient for
instance in the situations described in \cite{WK}; but when all particles
are measured, quantum mechanics sometimes predicts probabilities that cannot
be explained in terms of a pre-existing phase, and reveal a more fundamental
quantum character of this physical quantity. This creates the possibility of
new $N-$body violations of local realism with the use of Fock-state
condensates.

We thank W.D.\ Phillips for useful discussions.\ Laboratoire Kastler Brossel
is \textquotedblleft UMR 8552 du CNRS, de l'ENS, et de l'Universit\'{e}
Pierre et Marie Curie\textquotedblright .

\end{document}